\documentclass[traditabstract,letter]{aa} % for the abstract without structuration 
\usepackage{graphicx}
\usepackage{txfonts}
\usepackage{natbib}
\newcommand{\simless}{\mathbin{\lower 3pt\hbox
      {$\rlap{\raise 5pt\hbox{$\char'074$}}\mathchar"7218$}}} %< or of order
\newcommand{\simgreat}{\mathbin{\lower 3pt\hbox
     {$\rlap{\raise 5pt\hbox{$\char'076$}}\mathchar"7218$}}} %> or of order

\begin{document}
\title{Fomalhaut debris disk emission at 7 millimeters: \\ 
constraints on the collisional models of planetesimals}

\authorrunning{L. Ricci et al.}
\titlerunning{Fomalhaut Debris Disk Emission at 7~mm}

%   \subtitle{I. Overviewing the $\kappa$-mechanism}

   \author{L. Ricci\inst{1,2}
          \and
          L. Testi\inst{2,3}
          \and
          S.T. Maddison\inst{4}
          \and
          D.J. Wilner\inst{5}
          }

   \institute{ Division of Physics, Mathematics and Astronomy, California Institute of Technology, MC 249-17, Pasadena, 
CA 91125, USA\\
              \email{lricci@astro.caltech.edu}
         \and
             European Southern Observatory, Karl-Schwarzschild-Strasse 2, 85748 Garching, Germany
         \and
             INAF - Osservatorio Astroﬁsico di Arcetri, Largo Fermi 5, 50125 Firenze, Italy
         \and
         Centre of Astrophysics \& Supercomputing, Swinburne University of Technology, Hawthorn, VIC 3122, Australia    
	 \and
             Harvard-Smithsonian Center for Astrophysics, 60 Garden Street, Cambridge, MA 02138, USA
	     }

   \date{Received; accepted}

% \abstract{}{}{}{}{} 
% 5 {} token are mandatory

   \abstract{We present new spatially resolved observations of the dust thermal emission at 7~mm from the Fomalhaut debris disk obtained with the Australia Telescope Compact Array. These observations provide the longest wavelength detection of the Fomalhaut debris disk to date. We combined the new data to literature sub-mm data to investigate the spectral index of the dust thermal emission in the sub-millimeter and constrained the $q$-slope of the power-law grain size distribution. We derived a value for $q = 3.48 \pm 0.14$ for grains with sizes around 1 mm. This is consistent with the classical prediction for a collisional cascade at the steady-state. The same value cannot be explained by more recent collisional models of planetesimals in which either the velocity distribution of the large bodies or their tensile strength is a strong function of the body size.}
     

   \maketitle
%
%________________________________________________________________

\section{Introduction}

Dusty disks around nearby main-sequence stars have been known for almost three decades (Aumann et al.~1984). 
These are typically called debris disks, because the detected dust is thought to be produced by collisions between km-sized planetesimals, leftovers of the planetary formation process (see Wyatt~2008).

The highly destructive mutual encounters between planetesimals, triggered either by dynamical interaction with one or more planets in the system (Mustill \& Wyatt~2009) or by self-stirring (Wyatt~2008, and references therein), produce objects down to very small sub-$\mu$m dust grains.
The size distribution of the solids produced by these collisions provides information on some of the main properties of the invisible planetesimals. For example, collisional models of debris disks provide different predictions for the grain size distribution depending on the dynamical state (velocity dispersion, e.g. Pan \& Schlichting~2011) or physical conditions (e.g. tensile strength, Durda \& Dermott~1997) of the large bodies. 

Whereas the $\mu$m sized (and smaller) grains strongly interact with the stellar radiation via radiation pressure, the dynamics of $\sim$ 0.1 millimeter sized (and larger) pebbles are dominated by collisions and gravitational interactions. For this reason these pebbles remain located with the planetesimals (e.g. Wyatt~2005, Corder et al.~2009, Wilner et al.~2011). The spectral energy distribution (SED) of debris disks in the millimeter range can be used to constrain the size distribution of these pebbles and provide a powerful tool to test the collisional models of planetesimals (e.g. Vandenbussche et al.~2010).

In this paper, we present new spatially resolved 7~mm observations of the Fomalhaut debris disk obtained with the Australia Telescope Compact Array (ATCA).     
The Fomalhaut main-sequence star ($\alpha$ PsA, spectral type A4) is surrounded by one of the closest known dusty debris disks \citep[distance of 7.66$\pm$0.04 pc from Hipparcos,][]{van07}. Its age \citep[$\sim$ 200 Myr,] []{dif04} greatly exceeds the typical timescale for the survival of pristine circumstellar dust grains ($\simless$ 5-10 Myr) because of interaction with gas in primordial disks (see Pinilla et al.~2012, and Williams \& Cieza~2011 for a recent review), so that a continuous replenishment of dust through collisions of planetesimals is needed.

The geometry of the Fomalhaut debris disk is characterized by a very sharp inner edge at a radial distance of about 133~AU from the central star; the geometric center of the disk is also offset by about 15~AU from Fomalhaut (Kalas~2005); both these features were soon recognized as likely signatures for one or more planets shaping the morphology of the disk. 
A few years later, a planetary mass object, Fomalhaut b, has been imaged by Kalas et al.~(2008) using the \textit{Hubble Space Telescope} Advanced Camera for Surveys coronograph. This disk is therefore a benchmark for investigating the interaction between a planetary system and its debris disk (Chiang et al.~2009).

The dust continuum emission of this source has 
been measured in the sub-mm at about 0.35 mm with the CSO/SHARC II telescope (Marsh et al. 2005), at 0.45 mm and 0.85 mm with JCMT/SCUBA (Holland et al. 1998, 2003), and at 1.3 mm with SEST (Chini et al. 1991). The fluxes measured in the sub-mm make the Fomalhaut system one of the brightest debris disks at these long wavelengths.
Therefore this is one of the very few cases where current technology allows an investigation of the SED to wavelengths longer than $\sim 1$~mm. 

Together with the ATCA observations of the $\beta$ Pic debris disk published in Wilson et al.~(2011), our new observations provide the longest wavelength detection of any debris disk to date.
This gives us the advantage of probing grains as large as about 1~cm, which are virtually unaffected by the stellar radiation. At the same time, the addition of a flux measurement at the long-wavelength end of the SED allows us to better probe the Rayleigh-Jeans tail of the spectrum, thus minimizing the effect of any possible bias introduced by modeling the dust temperatures from the observed data. The new 7~mm data also provide a better lever arm in wavelength that mitigates the impact of absolute calibration uncertainties on the measurement of the spectral index. We present an analysis of the long-wave emission of the Fomalhaut debris disk and discuss its impact for the theories of collisional systems of planetesimals.     

\vspace*{-5mm}

\section{Observations}

We observed the Fomalhaut debris disk at 7~mm with ATCA and its Compact Array Broadband Backend (CABB) digital filter bank (Wilson et al.~2011).
The observations were carried out on September 15, 16, and 17, 2011. The ATCA array was in the H75 configuration, which provides baselines between 31 and 89~m. 
To obtain the best sensitivity in the continuum we set the correlator to cover
the full 4-GHz CABB effective bandwidth. The specific mean wavelength of the observations is 6.66~mm. 

\begin{figure*}[t!]
   \centering
\includegraphics[angle=90,scale=0.57]{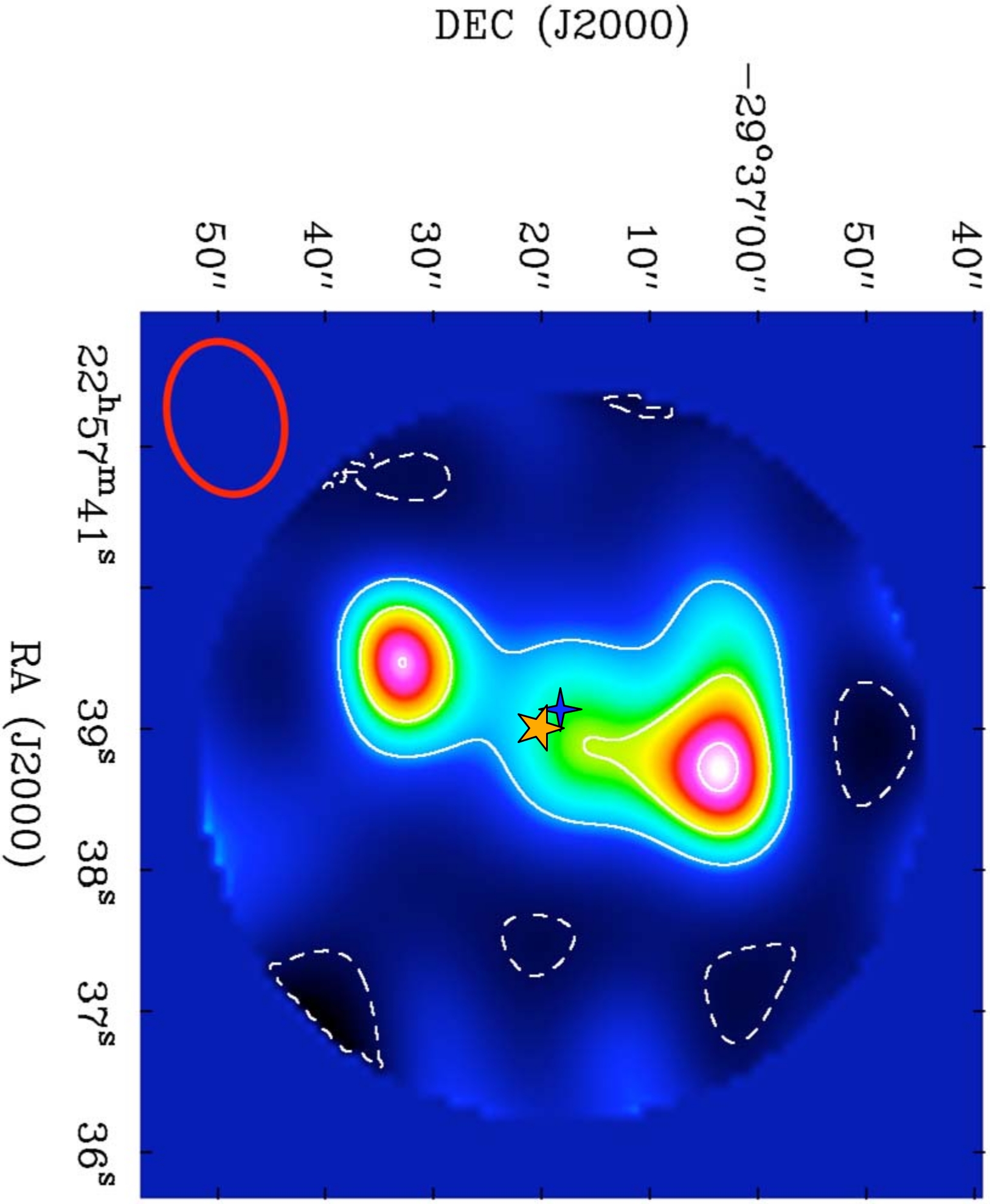}   
\caption{ATCA image at $\sim$~7~mm of the Fomalhaut debris disk corrected for the attenuation of the primary beam pattern up to the half power point ($\approx$~33$''$ from the pointing center). White contours are drawn at (-2, 2, 4, 6) $\times~\sigma$, where $1\sigma \approx 23$~$\mu$Jy is the rms noise level on the map. The orange star symbol identifies the location of the Fomalhaut star ($\alpha,\delta$)$=$(22:57:39.05,-29:37:20.05), whereas the disk center is indicated by the blue cross, whose size corresponds to the angular resolution of the ATCA observations ($\approx 2''$). Coordinates in the map have been corrected for the Fomalhaut proper motion \citep[329/-164 mas yr$^{-1}$, from][]{van07}, and are thus correct for the 2000.0 epoch. The red ellipse in the bottom left corner denotes the synthesized beam of the observations (FWHM $\approx$ 14.3$\times$10.7$''$, PA $\approx -76$~deg).}
              \label{fig:fomalhaut_atca}
   \end{figure*}

The raw visibilities for each day were calibrated and
edited with the MIRIAD software package.
The bandpass and absolute flux calibrations were made with 1921-293 and Uranus, respectively.
The complex gain calibration was performed on the calibrator
2255-282, and the Fomalhaut debris disk was typically observed
for 10 minutes followed by a 2 minute integration on the gain
calibrator.  
The total integration time on the Fomalhaut debris disk was about 12 hours.
On the three days of observations, the measured fluxes of the 2255-282 calibrator were found to be consistent within about 4\% from the mean value (3.61~Jy).
To account for the systematic effects that typically affect interferometric observations (e.g. baseline errors, pointing errors, etc.), we considered an uncertainty of about 15\% on
the calibrated flux at $\sim$~7 mm.
The 1$\sigma$ noise level in the final map is 23 $\mu$Jy.  
The angular resolution of our image is about 14.3 $\times$ 10.7$''$ (PA $\approx$ -76~deg), and the primary beam full-width-at-half-power (FWHP) is about about 65$''$ centered at the location of the Fomalhaut star (Figure~\ref{fig:fomalhaut_atca}).

\vspace*{-5mm}

\section{Results}

\subsection{ATCA map at 7mm}
\label{sec:ATCA_map}

Figure~\ref{fig:fomalhaut_atca} shows the new map of the Fomalhaut disk obtained with ATCA at 7 mm. The disk is clearly resolved into two main components, or lobes, northwest (NW) and southeast (SE) from the central star, respectively. These two lobes are detected at a signal-to-noise ratio of $\sim 6$. The emmision from the SE lobe is unresolved at the angular resolution of the ATCA observations. 

The two-lobe morphology seen in the ATCA map is very similar to the one observed at 0.85~mm by Holland et al.~(1998,2003) with the JCMT/SCUBA single-dish telescope at about the same angular resolution ($\approx 14''$).   
The position angle of the two lobes, evaluated at the peak of the surface brightness, is also nearly identical ($\approx 162, 159$ degrees from the ATCA and SCUBA maps, respectively). 
The angular resolution of our observations is not high enough to resolve the ringlike structure as observed by Marsh et al.~(2005) at 0.35~mm with the SHARC II camera at the CSO single-dish telescope.

In contrast to imaging from single-dish telescopes, interferometric imaging can filter out a fraction of the source emission. The largest angular scales that can be probed by interferometric observations are determined by the shortest spacings between any pair of antennas in the array.
The ATCA observations presented here are sensitive to emission at angular scales up to $\sim$ 50$''$. Since this is very close to the largest scales for the emission detected for the Fomalhaut disk by single-dish telescopes in the sub-mm, no significant filtering of emission from the disk is expected in our ATCA image. We verified this by taking a simple narrow ring model similar to that used by Marsh et al.~(2005) for the Fomalhaut disk; we simulated interferometric observations with the same coverage of the ($u,v$)-plane as for our ATCA observations, and we found that more than $99\%$ of the flux from the model was recovered by the simulated observations. 

In Figure~\ref{fig:fomalhaut_atca} the blue cross and orange star symbols show the locations of the disk center (the halfway point between the peaks of the NW and SE lobes) and of the star, respectively; the discrepancy between the positions of the disk center and of the star, i.e. about $2''$, or 15~AU, is consistent with that found by Kalas et al.~(2005) with HST. The fact that this same discrepancy is also seen at millimeter wavelengths confirms that the asymmetry involves the planetesimal ring, rather than just the smallest $\mu$m sized grains seen in the optical.  

The NW lobe appears to be more extended, showing two possible asymmetric structures toward east and south. These possible asymmetries are observed at low signal-to-noise ratios ($\approx 2-4$), and more sensitive observations are needed to determine the real nature of these apparent structures. The 4$\sigma$-contour in the southern asymmetry of the NW lobe extends to about 4$''$ from the location of the Fomalhaut star (see Figure~\ref{fig:fomalhaut_atca}); this emission, if real, might be caused by an ionized gas component (e.g. from a stellar wind or stellar corona) rather than by dust thermal emission from the debris disk; sensitive observations at centimeter wavelengths are needed to measure the spectral index of this emission and determine its nature. 

The integrated flux, including the central emission, is about $400 \pm 64$~$\mu$Jy (the uncertainty accounts for the absolute flux uncertainty), whereas the central emission is 92~$\mu$Jy, or 23\% of the total emission. The possible effect of the uncertainty on the naure of the central emission on our analysis will be discussed in Section~\ref{section:map}.

\vspace*{-4mm}

\subsection{Millimeter spectral index and slope of the grain size distribution}
\label{section:map}

The angular resolution of the new ATCA observations does not allow a detailed investigation of the morphology of the Fomalhaut debris disk. At the same time, the combination of the integrated flux derived at 7~mm with the sub-mm literature data allows us to probe the Rayleigh-Jeans tail of the dust thermal emission and derive information on the size distribution of the emitting grains.

Thermal dust emission from debris disks is optically thin, and therefore the flux density $F_{\nu} \propto B_{\nu}(T_{\rm{dust}}) \times \kappa_{\nu} \times M_{\rm{dust}}/d^2$, where $B_{\nu}(T_{\rm{dust}})$ is the Planck function at the characteristic dust temperature $T_{\rm{dust}}$, $\kappa_{\nu}$ is the dust opacity coefficient, $M_{\rm{dust}}$ is the dust mass, and $d$ is the distance. At long (sub-)millimeter wavelengths the dust opacity $\kappa_{\nu} \propto \nu^{\beta}$, and the Planck function $B_{\nu}(T_{\rm{dust}}) \propto \nu^{\alpha_{\rm{Pl}}}$ with $\alpha_{\rm{Pl}} \approx 2$. Hence, $\alpha_{\rm{mm}} \approx \alpha_{\rm{Pl}} + \beta$, where $\alpha_{\rm{mm}}$ is the (sub-)mm spectral index of the millimeter SED ($F_{\nu} \propto \nu^{\alpha_{\rm{mm}}}$). 

The $\beta$-parameter is related to the slope $q$ of the power-law differential size distribution of the dust grains ($dn(a) \propto a^{-q}~da$). Draine~(2006) derived the analytical formula $\beta \approx (q-3)\cdot\beta_{s}$, where $\beta_s$ is the dust opacity spectral index in the small particle limit. This is valid for $3 < q < 4$, and as long as the size distribution of grains is a single power-law over a broad enough interval of grain sizes around the observed wavelengths, i.e. $a_{\rm{min}} << 1$~mm, $a_{\rm{max}} >> 1$~mm (Draine~2006). 
This relation has been found to be very accurate for different models of dust considered in the literature to interpret the mm-wave emission of young circumstellar disks (see e.g. D'Alessio et al.~2001, Ricci et al.~2010a,b,2011).  
Combining the relations written above, one can obtain a simple relation linking the slope $q$ of the grain size distribution to the spectral indices $\alpha_{\rm{mm}}$, $\alpha_{\rm{Pl}}$, and $\beta_{s}$:

\begin{equation}
 q=\frac{\alpha_{\rm{mm}}-\alpha_{\rm{Pl}}}{\beta_s}+3,
\label{eq:1}
\end{equation}
which is valid for $3 < q < 4$. 

Assuming that particles in debris disks are composed of similar materials as interstellar grains and as are expected in primordial disks, the dust opacity spectral index in the small particle limit $\beta_s = 1.8 \pm 0.2$, where the uncertainty accounts for the different $\beta_s$-values measured in different diffuse interstellar clouds and in dense molecular clouds (see Draine~2006 and references therein).
The spectral index of the Planck function $\alpha_{\rm{Pl}}$ depends on the temperature of the emitting dust, which is equal to 2 in the Rayleigh-Jeans limit. Holland et al.~(2003) derived an interval of $\approx 40-50$~K for the temperature of the large ($\simgreat 1$~mm) grains observed in the millimeter, by fitting infrared and sub-mm data. This is consistent with the best-fit value of $42$~K derived by Marsh et al.~(2005). At these temperatures, the spectral index of the Planck function, calculated between 0.35 and 6.66~mm, is $\alpha_{\rm{Pl}} = 1.84 \pm 0.02$. 
Once the $\beta_s$ and $\alpha_{\rm{Pl}}$ parameters are estimated, Equation~\ref{eq:1} allows us to constrain the slope $q$ of the grain size distribution by measuring the spectral index $\alpha_{\rm{mm}}$ of the (sub-)mm SED.

   \begin{figure}[t!]
   \centering
\includegraphics[scale=0.5]{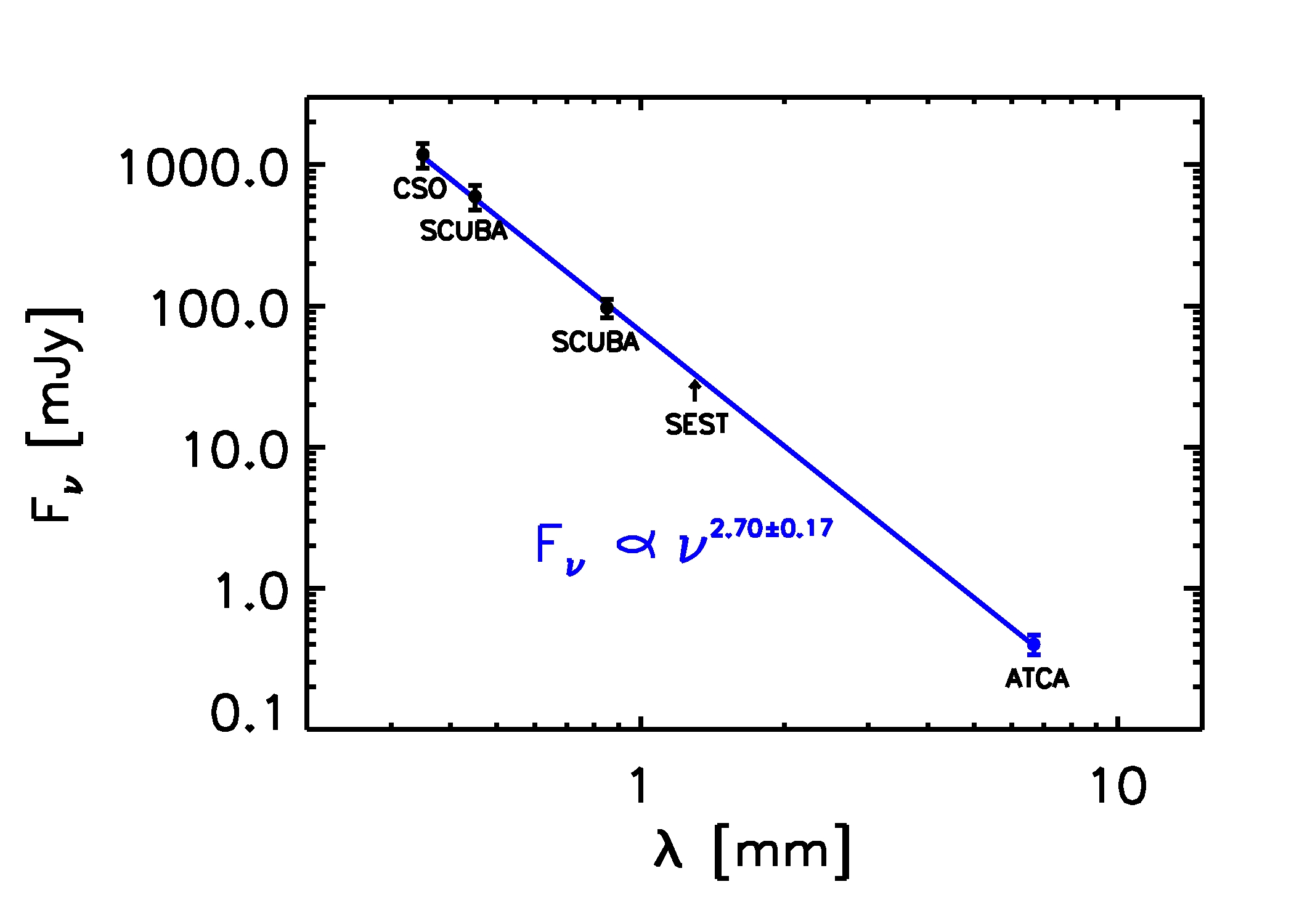}   
\caption{Sub-millimeter SED of the Fomalhaut debris disk. Data at 0.35, 0.45, 0.85 are from Marsh et al.~(2005) and Holland et al.~(2003); the plotted uncertainties for these data are $20\%$, $20\%$, and $15\%$ of the measured flux, respectively; the data point at 1.3~mm from Chini et al.~(1991) is taken as a lower limit and is not considered for the fit (Section~\ref{section:map}); the blue data-point indicates the integrated flux density at 6.66~mm from this paper. The value of the spectral index measured between 0.35 and 6.66~mm is also shown in blue.}
              \label{fig:fomalhaut_SED}
   \end{figure}

Figure~\ref{fig:fomalhaut_SED} shows the SED of the Fomalhaut debris disk between about 0.35~mm and 6.66~mm. Here we consider the flux measured at 1.3~mm by Chini et al.~(1991) only as a lower limit, because their SEST observations have been performed with a single beam (FWHP $\approx 24''$) pointing at the stellar position. Consequently, a significant fraction of the disk emission has been missed. Gaspar et al.~(2011) instead included the Chini et al.~(1991) detection in their analysis which explains the relatively high value of the spectral index they obtained, between 0.35 and 1.3~mm ($\approx 3.1 \pm 0.3$). In Figure~\ref{fig:fomalhaut_SED} we also report the measured value of the spectral index $\alpha_{\rm{0.35-6.66mm}} = 2.70 \pm 0.17$. 
Inserting this value in Equation~\ref{eq:1} yields a value for the slope of the grain size distribution $q = 3.48 \pm 0.14$, where the uncertainty is derived by propagating the uncertainties on the $\alpha_{\rm{mm}}, \alpha_{\rm{Pl}}, \beta_s$ parameters through Equation~\ref{eq:1}. The central emission at $\sim 7$~mm discussed above is associated with ionized gas and not with the debris disk, the constrained value of $q = 3.53 \pm 0.15$ would be fully consistent with the value reported above (the corresponding spectral index $\alpha_{\rm{0.35-6.66mm}} = 2.80 \pm 0.20$). The effect of the central emission on the derived value of $q$ is therefore negligible. 
We also checked that our result is not significantly biased by our choice of the interval in wavelength, e.g. due to deviations from the Rayleigh-Jeans regime especially at $\sim 0.35-0.45$~mm.
By repeating the same analysis in the 0.85-6.66~mm interval, we obtained $q = 3.43 \pm 0.17$, which is consistent with the value derived between 0.35 and 6.66~mm, although with a larger error because of the shorter lever arm in wavelength. 

\vspace*{-5mm}

\section{Discussion}
\label{sec:discussion}

Different collisional models of debris disks have provided predictions for the grain size distribution. In general, the destructive mutual encounters between planetesimals produce objects down to very small sub-$\mu$m grains; also, for most of the models, the predicted size distribution is satisfactorily approximated by a power-law with a slope $q$ assuming values between 3 and 4. Therefore, the assumptions used to derive Equation~\ref{eq:1} are satisfied and the constraints we derived in the last section for the $q$-slope of the Fomalhaut disk can be used to test the predictions of different models. 

This method allows one to determine the slope of the grain size distribution only in a relatively narrow interval of sizes between $\sim 0.1$~mm and $\sim 1$~cm, since these grains are the most efficient emitters in the sub-millimeter. Solids much larger than $\sim 1$~cm are not probed by our observations. But our method has the advantage of tracing grains that are virtually unaffected by the stellar radiation. Therefore, this provides more robust constraints for the grain size distribution than is currently possible by analyzing the infrared emission, which probes the smaller and more pressure radiation affected $\mu$m-sized particles.

The value of the $q$-index derived in the last section, i.e. $q = 3.48 \pm 0.14$, is consistent with the classical prediction of $q = 3.51$ for a collisional cascade at the steady-state, as a result of the collective dynamical interaction of particles caused by inelastic collisions and fragmentation (Dohnanyi~1969). 
The standard theoretical treatment of collisional cascades assumes a single constant tensile strength\footnote{The tensile strength of a body is defined as the minumum energy per unit mass required to disrupt it catastrophycally.} and velocity dispersion for all bodies regardless of size.
Both these quantities, and especially their possible variation with the size of the solids, are known to play a role in the predicted slope of the grain size distribution. 

Recently, Gaspar et al.~(2011) used numerical models of collision-dominated debris disks to investigate the effects of model parameters on the evolution of the grain size distribution. These authors showed that the slope of the tensile strength curve is the parameter that affects the slope of the grain size distribution the most (see also Durda \& Dermott~1997). In particular, the authors found that at the age of the Fomalhaut disk, values of $q \simgreat 3.82$ are obtained if the strength curve varies with the solids size as $a^{-1}$ or steeper. Adopting Gaspar et al.'s fiducial model for the Fomalhaut disk, our observations rule out such steep tensile strength curves at the $\simgreat 95\%$ confidence level. This discrepancy can be partially alleviated if other degenerate model parameters are simultaneously varied, but our observations always tend to favor shallower strength curves. In particular, the slope $q$ obtained by our analysis is consistent with the prediction of $q = 3.65$ of the Gaspar et al.~(2011) fiducial model, where they adopted a tensile strength curve varying with body size as $a^{-0.38}$.    

Pan \& Schlichting~(2011) have recently extended the classical treatment of collisional cascades (Donhanyi~1969) by relaxing the hypothesis of a single constant velocity dispersion, and solving self-consistently for the body size and size-dependent velocity distributions at the steady-state. They derived a velocity distribution $v(a) \propto a^{0.5}$ and a correspondingly very steep slope of the body's size-distribution $q = 4$ in the sizes of interest for the comparison with our results for the Fomalhaut disk. This predicted value of $q$ is ruled out by our analysis at the $99\%$ confidence level. Our analysis therefore suggests that the variation in body size of the velocity distribution of planetesimals in the Fomalhaut debris disk is weaker than predicted by the model of Pan \& Schlichting~(2011). This might be explained by the gravitational stirring expected in the Fomalhaut planetesimal belt caused by the known massive exo-planet (Fomalhaut b) detected close to the debris disk (Mustill \& Wyatt~2009).

The application of this method to other debris disks with and without massive exo-planets will be extremely interesting for investigating possible signs of interactions between planets and planetesimal belts. This will be made possible in the near future by the very high sensitivities and angular resolutions offered by ALMA and the EVLA at mm wavelengths.

\begin{acknowledgements}
  We thank the staff at Narrabri, ATNF.
\end{acknowledgements}

\vspace*{-8mm}

\bibliographystyle{aa}

\bibliography{ricci_fomalhaut}

\end{document}